\documentclass[superscriptaddress,reprint,amsmath,amssymb,aps,pre,floatfix]{revtex4-2}

\usepackage{graphicx}
\usepackage{dcolumn}
\usepackage{bm}
\usepackage{xcolor}
\usepackage[colorlinks = true,
             linkcolor = blue,
             urlcolor  = blue,
             citecolor = blue,
             anchorcolor = blue]{hyperref}

\begin{document}

\title{Vibrational resonance in a one-dimensional dissipative Bose-Josephson junction}

\author{Abhik Kumar Saha\,\,\href{https://orcid.org/0000-0001-8168-6742}
{\includegraphics[scale=0.05]{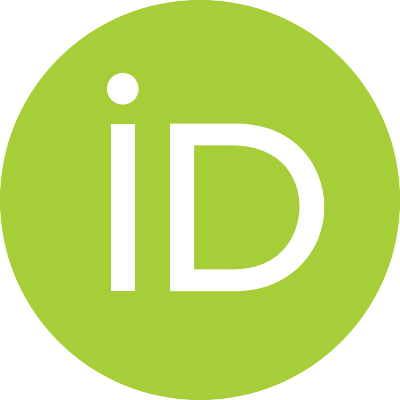}}}
\email{saha.abhikkumar.3k@kyoto-u.ac.jp}
\affiliation{Department of Physics, Kyoto University, Kitashirakawa Oiwakecho, Sakyo-ku, Kyoto 606-8502, Japan}
\date{\today}

\begin{abstract}
We investigate the linear and nonlinear response of a one-dimensional dissipative Bose-Josephson junction subjected simultaneously to a weak low-frequency probe and a rapidly oscillating high-frequency external drive. Starting from the dissipative two-mode Bose-Josephson equations, we derive an effective higher-order nonlinear equation for the population imbalance by retaining the leading nonlinear correction. Using time-scale separation and perturbative analysis, we obtain analytical expressions for both the linear response at the fundamental frequency and the nonlinear response at the second harmonic. We show that the high-frequency modulation modifies the effective potential landscape and dynamically breaks the symmetry around the stationary state, giving rise to a finite second-harmonic response that is absent without the rapidly oscillating field. Both the linear and nonlinear response amplitudes exhibit resonance-like enhancement for optimal values of the high-frequency driving strength. We further analyze the dependence of the resonance characteristics on interaction strength, dissipation, and driving parameters in both the zero-phase and $\pi$-phase modes and compare the analytical predictions with direct numerical simulations. Our results demonstrate a controllable mechanism toward realizing linear and nonlinear vibrational resonance in a one-dimensional dissipative Bose-Josephson junction and open new possibilities for controlling collective dynamics in driven ultracold bosonic systems.
\end{abstract}

\maketitle
\section{Introduction}
The Bose-Josephson junction (BJJ), realized in weakly coupled Bose-Einstein condensates confined in double-well (DW) potentials, provides a paradigmatic platform for studying coherent many-body quantum dynamics \cite{Bloch2008,Lewenstein2007,Gati2007,Schurer:2016,Haldar:2019,Bhowmik:2020,Boukobza:2010,Betz:2011,Hofferberth2007}, nonlinear tunneling phenomena \cite{shenoy:1999,josephson:exp,josephson:exp1,Josephson:dissipation,josephson:exp2,josephson:exp3,josephson:exp4,Milburn:1997,Zapata:1998,Raghavan:1999,Abad:2015,Levy:2007,Smerzi:2003}, and macroscopic quantum coherence \cite{Sakmann:2009,Ji:2022,Qiao2023,saha:2026_bjj}. Early theoretical and experimental studies established the existence of Josephson oscillations \cite{josephson:exp,josephson:exp1,josephson:exp2,shenoy:1999,Raghavan:1999,Milburn:1997}, $\pi$-phase oscillations \cite{Raghavan:1999,shenoy:1999,Albeiz:2005}, and macroscopic quantum self-trapping (MQST) \cite{Albeiz:2005,Gati2007,Pigneur2018,Pigneur:2018,Saha2021,Abad:2015}, demonstrating the rich interplay between tunneling and interatomic interactions in  bosonic systems. Owing to its high controllability, the BJJ has also emerged as a quintessential system for exploring non-equilibrium dynamics under external perturbations \cite{Saha2020,Josephson:dissipation,Juan:2019,Eckardt2005,LeBlanc:2011,Salmond:2002,Zhu:2021,Teichmann:2009}, dissipation \cite{STEFANATOS20192370,Mennemann:2021,Yuri:2021,Xhani:2022,Xhani2020dynamical,Ji:2022,Burchianti:2017,josephson:exp4,Josephson:dissipation,Bidasyuk:2016,saha:2026_bjj}, periodic driving \cite{Lin:2024_periodic,Boukobza:2010_periodic,Lin:2025_periodic,Gertjerenken:2014_periodic} and with analogous studies also extended to fermionic superfluid atomic systems \cite{josephson:fermi,Josephson-fermi-2:2020,Zaccanti:2019,Burchianti:2018,Xhani:2020,Kwon:2020}.   

In the mean-field description, the dynamics of BJJ is governed by coupled nonlinear equations for the population imbalance and relative phase between two condensates \cite{shenoy:1999,Raghavan:1999,Gati2007,Saha2021,josephson:exp,josephson:exp1,Milburn:1997,Albeiz:2005}. While the ideal BJJ dynamics is non-dissipative, several recent experiments and theoretical studies have demonstrated the existence of dissipative BJJ \cite{Xhani:2022,Xhani2020dynamical,josephson:exp4,Josephson:dissipation,STEFANATOS20192370,Mennemann:2021,Yuri:2021,Ji:2022,Burchianti:2017,Bidasyuk:2016,saha:2026_bjj}, where coupling to external environments, thermal fluctuations, or relaxation mechanisms leads to damping and equilibration dynamics \cite{Saha:2023,Mennemann:2021,Josephson:dissipation}. Dissipation introduces qualitatively new dynamical behavior, including relaxation toward phase-locked states \cite{Josephson:dissipation}, damping of coherent oscillations \cite{Xhani2020dynamical}, and modified nonlinear tunneling dynamics \cite{Binati:2021,Sinha:2019,Szirmai:2015}. These developments have generated considerable interest in understanding how external forcing and dissipation together influence the collective dynamics of interacting bosons. 

Another important aspect of driven nonlinear systems is the phenomenon of vibrational resonance \cite{Landa_2000,Yang_2024,Baltanas:2003,Abirami:2017,Jeyakumari:2009}, where the response of a nonlinear system to a weak low-frequency signal can be strongly enhanced by the presence of an additional high-frequency periodic drive. Vibrational resonance has been extensively investigated in a wide variety of classical nonlinear systems, including bistable systems \cite{ULLNER2003348,Gandhimathi_2006}, Duffing oscillators \cite{Jeevarathinam:2011,Ivan:2021,Jeyakumari,jeevarathinam:2015}, biological systems \cite{Yang_2024,Yu:2011,Rajasekar:2012}, quantum dynamics \cite{Sarkar:2021,Paul:2021,Sarkar:2019,Ghosh:2015,Das:2018} and nonlinear circuits \cite{JOTHIMURUGAN_2013,Abirami_2014}. The essential mechanism arises from the modification of the effective potential landscape induced by the rapidly oscillating field, which can optimize the response to a weak probe signal. In recent years, increasing attention has also been devoted to nonlinear and higher-harmonic responses generated through the interference of multiple time scales in periodically driven nonlinear systems \cite{Linsay1981,Gammaitoni1998,Nayfeh1979}.  

In the context of BJJ, externally driven dynamics has already revealed several intriguing phenomena \cite{Eckardt2005,Jinasundera2006,Gertjerenken2014,Sakellari2004,Singh:2024} such as parametric oscillations \cite{Saha2020}, self-trapping resonances \cite{Saha2021}, and dissipative relaxation dynamics \cite{Xhani:2022,Xhani2020dynamical,josephson:exp4,Josephson:dissipation}. However, most existing studies of driven BJJ have primarily focused on the dynamics induced by a single periodic drive \cite{Eckardt2005,Saha2021,Singh:2024}. In particular, the role of a rapidly oscillating high-frequency field in generating linear and nonlinear harmonic response in dissipative BJJ remains largely unexplored. Moreover, although nonlinearities naturally arise in the Josephson dynamics through interaction effects, the influence of higher-order nonlinear corrections on the response properties of dissipative BJJ has not yet been systematically analyzed. An especially important unresolved issue concerns the emergence of nonlinear harmonic generation in symmetric BJJ. In a symmetric DW potential, the effective restoring force contains only odd powers of the population imbalance, preserving inversion symmetry and consequently forbidding even-harmonic generation. Therefore, in the absence of symmetry breaking, the response at the second harmonic frequency is expected to vanish. It is worthwhile to understand how external high-frequency modulation may effectively break this symmetry and generate finite nonlinear response in dissipative BJJ. Furthermore, the dependence of such nonlinear response on interaction strength, dissipation, and driving parameters has not been investigated analytically and numerically within a unified framework.

Motivated by these questions, we investigate in this work the linear and nonlinear response of a one-dimensional (1D) dissipative BJJ subjected simultaneously to a weak low-frequency signal and a rapidly oscillating high-frequency drive. Starting from the dissipative two-mode BJJ equations, we derive an effective higher-order nonlinear equation for the population imbalance by retaining the leading cubic nonlinear correction. Using the method of time-scale separation and perturbative analysis, we obtain analytical expressions for both the linear response amplitude at the fundamental frequency and the nonlinear response amplitude at the second harmonic frequency. Our analysis demonstrates that the high-frequency periodic modulation modifies the effective potential landscape and induces an effective symmetry breaking around the stationary state, thereby generating a finite second-harmonic response which is absent without the high frequency field. We further show that both the linear and nonlinear response amplitudes exhibit resonance-like enhancement for optimal values of the high-frequency driving strength. The resonance characteristics are analyzed in both the $\pi$-phase mode with positive interaction and the zero-phase mode with negative interaction. In addition, we investigate how the resonance position depends on the interaction parameter, high frequency drive, and low-frequency forcing amplitude. Analytical predictions are compared with direct numerical simulations of both the reduced higher-order nonlinear 1D dissipative BJJ model and the full dissipative BJJ equations, showing overall satisfactory agreement over a broad range of parameters. The present study therefore establishes a mechanism for controllable nonlinear harmonic generation and vibrational resonance in a 1D dissipative BJJ through external high-frequency modulation.

The paper is organized as follows. In Sec.~\ref{sec:2}, we introduced the dissipative BJJ and derive an effective higher-order nonlinear equation governing the population imbalance dynamics by retaining the leading nonlinear correction within the two-mode framework. In Sec.~\ref{sec:3}, we consider the effects of external time-dependent perturbations consisting of a weak low-frequency signal and a rapidly oscillating high-frequency field. We define the linear and nonlinear response functions. Using time-scale separation together with perturbative analysis, we derive analytical expressions for linear and nonlinear response amplitude.  In Sec.~\ref{sec:4}, we present and discuss the numerical simulations for both the reduced higher-order nonlinear 1D dissipative BJJ model and the full dissipative BJJ equations to validate the theoretical predictions. We analyze the resonance behavior in both the $\pi$-phase  and zero-phase modes and investigate the dependence of the resonance position on interaction strength and external driving parameters. Finally, the paper is concluded in Sec.~\ref{sec:5}.

\section{Dissipative Bose-Josephson junction: A higher order nonlinear model}\label{sec:2}
Usually the Josephson oscillations in a DW potential is non-dissipative. This implies the dynamics of the atom number imbalance and the relative phase remains undamped over time \cite{shenoy:1999,josephson:exp,josephson:exp1,Albeiz:2005}. However, in recent times, several studies have been reported a dissipative BJJ which is analogous to a pendulum with friction \cite{Mennemann:2021,Xhani:2022,Pigneur:2018}.

The dissipative BJJ \cite{Xhani:2022,Xhani2020dynamical,josephson:exp4,Josephson:dissipation,Pigneur2018,Marino:1999} is defined in terms of normalized atom number imbalance $z(t)$ and phase difference $\theta(t)$ in dimensionless unit: 
\begin{eqnarray}
\dot{z}(t)=-\sqrt{1-z^2(t)}\sin\theta(t)-\gamma z(t)
\label{eq:z}
\end{eqnarray}
\begin{eqnarray}
\dot{\theta}(t)=\Delta E+\frac{z(t)}{\sqrt{1-z^2(t)}}\cos\theta(t)+\Lambda_{0}z(t)
\label{eq:theta}
\end{eqnarray}
The atom number imbalance is defined as $z=\frac{N_L-N_R}{N_L+N_R}$. The conjugate variable is the relative phase defined by $\theta=\theta_R-\theta_L$. $N_{L(R)}$ and $\theta_{L(R)}$ is the number of atoms and the phase of the atoms in the left(right) well. $\Delta E$ represents the asymmetry between the two-well. $\Lambda_0(=\frac{NU}{2J})$ characterizes the many-body interaction parameter with $U$ being the on-site mean two-body interaction energy and $J$ is the tunneling amplitude between two sites of the DW. $\gamma$ represents the dissipation coefficient of the system. Eqs.~(\ref{eq:z}) and (\ref{eq:theta}) represent the dissipative BJJ equations in which the dissipative coefficient term can be modified by the linear contribution from $\dot{\theta}(t)$, which is usually studied in the standard dissipative BJJ \cite{Pigneur2018,Marino:1999}. In the absence of dissipation $(\gamma=0)$, Eqs. (\ref{eq:z}) and (\ref{eq:theta}) reduce to the standard BJJ equations \cite{Raghavan:1999,shenoy:1999}. The simplified form of the dissipative BJJ (Eqs. (\ref{eq:z}) and (\ref{eq:theta})) in terms of population imbalance $z$ and phase difference $\theta$ arises by considering two-mode model of a BJJ coupled to two-bosonic baths. The coupling between the system and bath modes can be considered as the usual linear system-bath coupling \cite{Saha2020} or nonlinear \cite{Saha:2023}, such that the excitation in the bath modes is not accompanied by energetic deexcitation of the system. While both couplings have applications in 1D dissipative BJJ, the first one used for the study of parametric oscillation \cite{Saha2020} and the latter one used for the study of phase diffusion and fluctuation properties \cite{Saha:2023} in dissipative BJJ.

To explore the effect of the leading order nonlinearlity in the dissipative BJJ (Eqs.~(\ref{eq:z}) and (\ref{eq:theta})) under the presence of time-dependent external perturbation, we proceed as follows: For small population imbalance and small phase difference around zero, one can approximate $z\approx 0$, $\theta\approx 0$, $\sqrt{1-z^2(t)}\approx 1-\frac{z^2}{2}+\mathcal{O}(z^4)$, $\sin\theta\approx\theta$, $\cos\theta\approx1$ and $\frac{z}{\sqrt{1-z^2}}\approx z+\frac{z^3}{2}+\mathcal{O}(z^5)$. Making use this approximation, Eqs. (\ref{eq:z}) and (\ref{eq:theta}) becomes
\begin{eqnarray}
\dot{z}(t)\approx-\left[1-\frac{z^2(t)}{2}\right]\theta(t)-\gamma z(t)
\label{eq:z-approx}
\end{eqnarray}
\begin{eqnarray}
\dot{\theta}(t)\approx \Delta E+\left[\Lambda_{0}+1\right]z+\frac{z^3}{2}    
\end{eqnarray}
Note that in the above equations, we have retained the leading order nonlinear terms up to third order in $z$. From Eq.~(\ref{eq:z-approx}), the approximate equation for the phase difference follows:
\begin{equation}
\theta(t)\approx\left[-\dot{z}(t)-\gamma z(t)-\frac{z^2}{2}\dot{z}(t)-\frac{\gamma z^3}{2}\right]    
\end{equation}
In the above equation, the main contribution is coming from the first two terms in the right hand side and since $\gamma$ is small so we neglect the last term and the mixed term. By differentiating Eq.~(\ref{eq:z}) with respect to time, we write
\begin{equation}
\ddot{z}(t)=\frac{z\dot{z}}{\sqrt{1-z^2}}\sin\theta-\sqrt{1-z^2}\cos\theta\dot{\theta}-\gamma \dot{z}
\label{eqn-zddot}
\end{equation}
Now applying the above approximation and retaining nonlinear terms up to third order in $z$, while substituting the values of $\theta$ and $\dot{\theta}$ in Eq.~(\ref{eqn-zddot}) and neglecting mixed terms of $z$ and $\dot{z}$ as well as their higher-order contributions, we obtain a simplified equation for the zero-phase mode:
\begin{equation}
\ddot{z}+\gamma \dot{z}+[\Lambda_0+1]z-\frac{\Lambda_0z^3}{2}=-\Delta E    
\end{equation}
For the $\pi$-phase mode, the above equation becomes
\begin{equation}
\ddot{z}+\gamma \dot{z}-[\Lambda_0-1]z+\frac{\Lambda_0z^3}{2}=\Delta E 
\end{equation}
For simplicity, we define $\alpha_0=\Lambda_0+1$, $\beta_0=-\Lambda_0/2$ and $\alpha_{\pi}=1-\Lambda_{0}$, $\beta_{\pi}=\Lambda_0/2$. With these definitions, the above two equations become:
\begin{eqnarray}
\ddot{z}+\gamma \dot{z}+\alpha_{0,\pi}z+\beta_{0,\pi}z^3=\mp\Delta E\label{eq:underdamped1}
\end{eqnarray}
It is important to note that the values of $\sqrt{\alpha_0}$ and $\sqrt{\alpha_{\pi}}$ denote the Josephson frequency at zero and $\pi$-phase mode in dimensionless unit \cite{Raghavan:1999,Saha2020,shenoy:1999}. To get stable solution of the above equations, it is worth noting that the coefficient of linear and non-linear population imbalance should follow $\alpha_{0,\pi}\textless0$ and $\beta_{0,\pi}\textgreater0$. This implies the negative values of the $\Lambda_0\textless-1$ in the zero-phase mode but positive values of $\Lambda_0\textgreater1$ in the $\pi$-phase mode. In absence of asymmetry ($\Delta E=0$) and neglecting non-linear order in the population imbalance the above equations becomes damped harmonic oscillator which is usually studied in the standard dissipative BJJ \cite{Saha2020,Marino:1999}. It is worth mentioning that for negative $\Lambda_0$ ($\Lambda_0\textless-1$), the small $z$ linearization around zero-phase mode gives the linearized frequency imaginary. It implies that $z=0$ equilibrium is unstable and the system undergoes a bifurcation. In other words the system moves away from symmetric $z=0$ state and settle to new asymmetric equilibria (non-zero population imbalance). This behavior corresponds to the onset of symmetry breaking and MQST in the two-mode BJJ with negative interaction \cite{Raghavan:1999,josephson:exp}. In particular, the conditions for $\Lambda_0$ in both zero and $\pi$-phase mode signifies the condition for the MQST for the respective cases by the inclusion of higher order nonlinear term. Upon closer look of the above equation (Eq.~(\ref{eq:underdamped1})) including the nonlinear terms in population imbalance also signifies that, both equations gives an effective potential description in the underdamped limit.

\section{Time-dependent perturbation and response amplitudes}\label{sec:3}
Now we make the asymmetry of the DW potential to be time dependent of the form $\Delta E(t)=c\cos(\omega t)+g\cos(\Omega t)$.  $c\cos(\omega t)$ is the low frequency input signal and $g\cos(\Omega t)$ corresponds to the rapidly oscillating periodic force with frequency $\Omega\gg\omega$. 
Such time-dependent periodic modulation of the energy bias is experimentally well motivated in BJJ systems. Periodically driven DW condensates have already been realized using dynamic modulation of the trapping potential, where the relative energy offset between the wells is controlled in time through external driving fields \cite{josephson:exp1,Eckardt2005,LeBlanc:2011,Salmond:2002,Tozzo:2005,Zhu:2021,Teichmann:2009}. Such setups have enabled the observation of photon-assisted tunneling \cite{Eckardt2005}, Shapiro-like resonances \cite{Eckardt2005,Teichmann:2009}, and interaction-induced nonlinear effects in driven BJJ systems \cite{LeBlanc:2011,Salmond:2002,Tozzo:2005,Zhu:2021}. The use of a superposition of slow and fast periodic drives therefore provides a natural extension of existing experimental protocols and offers an experimentally feasible route for exploring multiscale nonlinear response and vibrational resonance phenomena.  

\subsection{Linear and nonlinear response}
The linear response of the system is evaluated by calculating the sine and cosine components $B_s(\omega)$ and $B_c(\omega)$, respectively of the output signal or the state of the system denoted by $z(t)$ as follows:
\begin{equation}
B_s(\omega)=\frac{2}{nT}\int_{0}^{nT}z(t)\sin(\omega t)
\label{eq:10}
\end{equation}
\begin{equation}
B_c(\omega)=\frac{2}{nT}\int_{0}^{nT}z(t)\cos(\omega t)
\label{eq:11}
\end{equation}
where $T=\frac{2\pi}{\omega}$ with integer $n$. Solving Eq.~(\ref{eq:underdamped1}) in presence of time-dependent perturbation followed by extraction of its sine and cosine components yields the linear response function
\begin{equation}
Q_{\rm L}(\omega)=\sqrt{B_s^2(\omega)+B_c^2(\omega)}/c
\label{eq:12}
\end{equation}
and phase shift $\phi(\omega)=\tan^{-1}\left[\frac{B_s(\omega)}{B_c(\omega)}\right]$ of the response function with respect to input signal.

It is quite straightforward to extend the definition of linear response amplitude to nonlinear domain by calculating the sine and cosine components of $z(t)$ at the second harmonic frequency of the input signal,
\begin{equation}
B_s(2\omega)=\frac{2}{nT}\int_{0}^{nT}z(t)\sin(2\omega t) 
\label{eq:13}
\end{equation}
\begin{equation}
B_c(2\omega)=\frac{2}{nT}\int_{0}^{nT}z(t)\cos(2\omega t) 
\label{eq:14}
\end{equation}
where $T$ is now redefined as $T=\frac{\pi}{\omega}$. Similarly we define the phase shift 
$\phi(2\omega)=\tan^{-1}\left[\frac{B_s(2\omega)}{B_c(2\omega)}\right]$ of the response function at the second harmonic frequency. The nonlinear response amplitude at the second harmonic can be determined by calculating
\begin{equation}
Q_{\rm NL}(2\omega)=\sqrt{B_s^2(2\omega)+B_c^2(2\omega)}/c^2 
\label{eq:15}
\end{equation}

We now emphasize a pertinent point at this stage. Due to the inversion symmetry $z\rightarrow-z$ of the effective potential associated with Eq.~(\ref{eq:underdamped1}), the system does not exhibit a response at the second harmonic $2\omega$ in the absence of the high-frequency driving field ($g=0$). Indeed, the restoring force contains only odd powers of $z$, leading to a symmetric potential landscape and consequently forbidding even-harmonic generation. However, in the presence of the additional high-frequency drive, nonlinear mixing between the low- and high-frequency components can induce a finite response at $2\omega$. In the following, we show that an optimal range of the high-frequency drive amplitude $g$ leads to the emergence and enhancement of the nonlinear second-harmonic response.

\subsection{Effective dynamics via time-scale separation and perturbation analysis}
To obtain an appropriate theoretical estimate of the linear $Q_{\rm L}(\omega)$ and nonlinear $Q_{\rm NL}(2\omega)$ response amplitude, we identify, following standard technique, two time scales of the dynamics and seek for a solution of the form
\begin{equation}
z=X(t,\omega t)+\psi(t,\Omega t)    
\end{equation}
Here $X(t,\omega t)$ and $\psi(t,\Omega t)$ correspond to the slow and the fast motion components, respectively. $\psi$ is $2\pi$ periodic and therefore has zero mean as 
\begin{equation}
\langle \psi(t,\tau)\rangle=\frac{1}{2\pi}\int_{0}^{2\pi} \psi(t,\tau)d\tau   
\end{equation}
Here $\tau=\Omega t$ refers to the fast time scale. 

By putting the time-dependent asymmetry, Eq.~(\ref{eq:underdamped1}) becomes 
\begin{eqnarray}
\ddot{z}+\gamma\dot{z}+\alpha_{0,\pi}z+\beta_{0,\pi}z^3=\mp c \cos(\omega t)\mp g\cos(\Omega t) 
\label{eq-underdamped3}
\end{eqnarray}
Making use of the decomposition of Eq.~(\ref{eq-underdamped3}) and averaging over the fast time scale, we obtain, for slowly moving variable $X$,
\begin{eqnarray}
\ddot{X}&+&\gamma\dot{X}+\alpha_{0,\pi}X+\beta_{0,\pi}X^3\nonumber\\&+&3\beta_{0,\pi}X^2\langle\psi\rangle+3\beta_{0,\pi}X\langle\psi^2\rangle=\mp c\cos(\omega t)    
\end{eqnarray}
and for the fast motion,
\begin{eqnarray}
\ddot{\psi}&+&\gamma\dot{\psi}+\alpha_{0,\pi}\psi+\beta_{0,\pi}\psi^3+3\beta_{0,\pi}X^2\left[\psi-\langle \psi\rangle\right]\nonumber\\&+&3\beta_{0,\pi}X\left[\psi^2-\langle \psi^2\rangle\right]=\mp g\cos(\Omega t) 
\label{eq:psi_underdamped}
\end{eqnarray}
$\psi$ being a rapidly changing field. We assume further $\dot{\psi}\gg\psi,\psi^2,\psi^3$. The dynamics of $\psi$ can be written down as 
\begin{eqnarray}
\dot{\psi}=\mp\frac{g\gamma\cos(\Omega t)}{\gamma^2+\Omega^2}\mp \frac{g\Omega\sin(\Omega t)}{\gamma^2+\Omega^2}
\end{eqnarray}
so that its solution yields 
\begin{eqnarray}
\dot{\psi}=\mp\frac{g\gamma\sin(\Omega t)}{\Omega(\gamma^2+\Omega^2)}\pm \frac{g\cos(\Omega t)}{\gamma^2+\Omega^2}
\end{eqnarray}
and as follows, $\langle \psi\rangle=\langle \psi^3\rangle=0$ and \begin{eqnarray}
\langle \psi^2\rangle=\frac{g^2}{2}\left[ \frac{1+\gamma^2/\Omega^2}{(\gamma^2+\Omega^2)^2}\right]    
\end{eqnarray}
The motion of the slow component then becomes
\begin{eqnarray}
\ddot{X}+\gamma\dot{X}+ A(g)X+\beta_{0,\pi}X^3=\mp c\cos(\omega t),
\label{eqn-cX}
\end{eqnarray}
where, $A(g)=\alpha_{0,\pi}+\frac{3\beta_{0,\pi}g^2}{2}\left[ \frac{1+\gamma^2/\Omega^2}{(\gamma^2+\Omega^2)^2}\right]$. A look into Eq.~(\ref{eqn-cX}) reveals that the effect of high frequency oscillation due to rapidly changing force $g\cos(\Omega t)$ in Eq.~(\ref{eq:psi_underdamped}) is included in $A(g)$. Eq.~(\ref{eqn-cX}) describes the under-damped dynamics of a particle in a DW potential where the potential is modified by the strength and frequency of the high frequency field.

The time-independent steady state solution $X_s$ can be obtained by setting simultaneously $\dot{X}=0$ and $\ddot{X}=0$, and the forcing has zero time dependence, then Eq.~(\ref{eqn-cX}) becomes
\begin{eqnarray}
A(g)X_s+\beta_{0,\pi}X_s^3=0
\label{eq:ag}
\end{eqnarray}
with solutions $X_s=0$, and $X_s^2=-\frac{A(g)}{\beta_{0,\pi}}$. Since $\beta_{0,\pi}\textgreater0$, so the non-zero steady state becomes real only when $A(g)\textless0$ which gives a critical value of the fast oscillating amplitude $g_{c}$ in the under-damped limit. To make $A(g)\textless0$, we use 
$g\textless g_{c}\equiv\sqrt{\frac{2\Omega^2(\gamma^2+\Omega^2)(-\alpha_{0,\pi})}{3\beta_{0,\pi}}}$ with $\alpha_{0,\pi}\textless0$. Thus, the fixed points can be controlled by the ratio $(\frac{g}{\Omega})$ of the rapidly varying field. The evolution of the dynamics around the steady state can be determined by introducing the perturbation variable $Y=X-X_S$, so that Eq.~(\ref{eqn-cX}) becomes
\begin{eqnarray}
\ddot{Y}&+&\gamma\dot{Y}-2YA(g)\nonumber\\&+&3\beta_{0,\pi}X_sY^2+\beta_{0,\pi}Y^3=\mp c\cos(\omega t)    
\label{eq-per}
\end{eqnarray}
We now emphasize two pertinent points. First, we do not consider here the perturbation to be small. This allows us to calculate the nonlinear response of the system. Second, a comparison between Eqs.~(\ref{eqn-cX}) and (\ref{eq-per}) shows that the center of symmetry of the potential is destroyed in the effective dynamics of $Y$ due to the appearance of the term $3\beta_{0,\pi} X_SY^2$, which assures a nonzero value of the steady state $X_S$. The loss of symmetry arises due to the interference of the high frequency temporal oscillation and as we show here, allows us to calculate the response of the system at the second harmonic of the low frequency probe in addition to the usual linear response function. An inspection of Eq.~(\ref{eq-per}) shows that the term $3\beta_{0,\pi} X_SY^2$ gives rise to a component oscillating at $2\omega$.

\subsection{Analytical expressions for linear and nonlinear response amplitude}
We now calculate the response of the system at the first and second harmonic of the low frequency probe (to the low frequency field). Therefore, we assume the solution for $Y(t)$ in the form
\begin{align}
Y(t) &= k_1\cos(\omega t) + k_2\sin(\omega t) \nonumber \\
     &\quad + k_3\cos(2\omega t) + k_4\sin(2\omega t)
\label{eq:sol}
\end{align}
Substituting the last expression into Eq.~(\ref{eq-per}) and equating the coefficients of $\cos(\omega t)$, $\sin(\omega t)$, $\cos(2\omega t)$ and $\sin(2\omega t)$, we finally obtain 
\begin{eqnarray}
Y(t)=Y_{\rm L}(t)+Y_{\rm NL}(t)    
\end{eqnarray}
where $Y_{L}=K\cos(\omega t-\delta)$ and $Y_{\rm NL}=L\cos(2\omega t-\eta)$, and
\begin{eqnarray}
K=\sqrt{k_1^2+k_2^2}=\frac{c}{\sqrt{\{\omega^2+2A(g)\}^2+\omega^2\gamma^2}}    
\end{eqnarray}
The linear response amplitude is given by
\begin{eqnarray}
Q_{\rm L}(\omega)=\frac{K}{c}=\frac{1}{\sqrt{\{\omega^2+2A(g)\}^2+\omega^2\gamma^2}}
\label{eq:ana_linearresponse}
\end{eqnarray}
The nonlinear response, $Y_{\rm NL}(t)$, on the other hand assumes a similar form $Y_{\rm NL}=L\cos(2\omega t-\eta)$, where
\begin{equation}
\begin{aligned}
L &= \sqrt{k_3^2 + k_4^2} \\
  &= \frac{(3\beta_{0,\pi}X_S/2)c^2}
  {\left[\{\omega^2 + 2A(g)\}^2 + \omega^2 \gamma^2\right]
  \sqrt{\left(4\omega^2 + 2A(g)\right)^2 + (2\omega\gamma)^2}}
\end{aligned}
\end{equation}
The nonlinear response amplitude at the second harmonic is therefore given by
\begin{equation}
Q_{\rm NL}(2\omega)
= \frac{L}{c^2}
= \frac{(3\beta_{0,\pi}X_S/2)}
{D(\omega)\,
\sqrt{\left(4\omega^2 + 2A(g)\right)^2 + (2\omega\gamma)^2}}
\label{eq:ana_nonlinearresponse}
\end{equation}
where $D(\omega) = \left[\{\omega^2 + 2A(g)\}^2 + \omega^2 \gamma^2\right]$.
The nature of resonance at the first (Eq.~(\ref{eq:ana_linearresponse})) and second (Eq.~(\ref{eq:ana_nonlinearresponse})) harmonic depends on the structure of $A(g)$ which is determined by the condition (Eq.~(\ref{eq:ag})). The range of values of $g$ for which $A(g)$ is negative determines the resonances. We explore this issue in more detail in the next section.

\section{Results and Discussion}\label{sec:4}
In order to confirm our theoretical observations, we have carried out numerical simulations for several values of parameters. We have solved Eq.~(\ref{eq-underdamped3}) with positive ($\pi$-phase mode) and negative (zero-phase mode) interaction parameter numerically to calculate the linear $Q_{\rm L}(\omega)$ and nonlinear $Q_{\rm NL}(2\omega)$ response of the system using Eqs. (\ref{eq:10}), (\ref{eq:11}), (\ref{eq:12}), (\ref{eq:13}), (\ref{eq:14}) and (\ref{eq:15})  and compared the results with our analytical estimate (Eq.~(\ref{eq:ana_linearresponse})) of the linear response and (Eq.~(\ref{eq:ana_nonlinearresponse})) of the nonlinear response. To validate our perturbation technique, we have also carried out direct simulation of Eqs.~(\ref{eq:z}) and (\ref{eq:theta}), i.e simulating full 1D dissipative BJJ without any approximation using Eqs. (\ref{eq:10}), (\ref{eq:11}), (\ref{eq:12}), (\ref{eq:13}), (\ref{eq:14}) and (\ref{eq:15}). In the following, we present the results for the $\pi$-phase and zero-phase mode of the 1D dissipative BJJ. This ensures the variation of the linear and nonlinear response amplitude for 1D dissipative BJJ in presence of both positive and negative interaction parameter.

\subsection{$\pi$-phase mode}
We first consider the positive onsite interaction parameter with initial values of population imbalance $z_{0}$ and phase difference $\theta_{0}$ is $(z_{0},\theta_{0})=(0.1,\pi)$ for $\pi$-phase mode. The chosen parameter set is $\Lambda_0=1.2$ and $\gamma=2.4$, i.e $\Lambda_0/\gamma=0.5$.
For our numerical simulation, we fixed $\omega=0.1$ and $c=0.05$. The evolution of the system is followed for over a time $t=10T$ for calculating linear response and $t=20T$ for calculating nonlinear response of the system. To calculate the sine and cosine components of $B_s(\omega)$, $B_s(2\omega)$, $B_{c}(\omega)$, and $B_{c}(2\omega)$, we consider the last $5$ ($n=5$) and $10$ ($n=10$) periods of time for linear and nonlinear response of the system. 

\begin{figure}[htbp]
\centering
\includegraphics[width=0.48\textwidth]{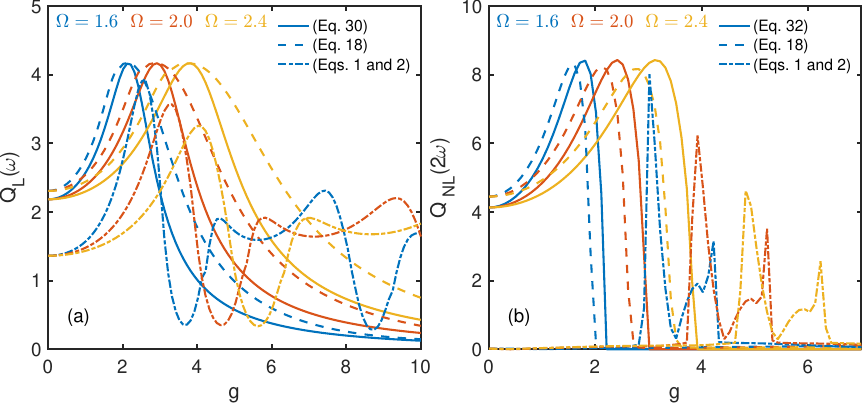}
\caption{(Color Online) (a) Linear response amplitude $Q_{\rm L}(\omega)$ and (b) nonlinear response amplitude $Q_{\rm NL}(2\omega)$ of the 1D dissipative BJJ for $\Lambda_{0}/\gamma = 0.5$, with a positive interaction parameter in the $\pi$-phase mode, in response to a low-frequency signal $c\cos(\omega t)$ under the influence of a high-frequency drive $g\cos(\Omega t)$. Results are shown for three different values of $\Omega = 1.6$ (blue), $2.0$ (orange), and $2.4$ (yellow), with fixed parameters $c = 0.05$ and $\omega = 0.1$. The solid lines represent the analytical results for $Q_{\rm L}$ (Eq.~(\ref{eq:ana_linearresponse})) and $Q_{\rm NL}$ (Eq.~(\ref{eq:ana_nonlinearresponse})). The dashed lines correspond to numerical simulations of $Q_{\rm L}(\omega)$ and $Q_{\rm NL}(2\omega)$ using Eq.~(\ref{eq-underdamped3}). The dash-dotted lines represent numerical simulations using Eqs.~(\ref{eq:z}) and (\ref{eq:theta}). For better visibility of the resonant behavior as a function of $g$, the results from the full direct simulation of the 1D dissipative BJJ (Eqs.~ (\ref{eq:z}) and (\ref{eq:theta})) are multiplied by factors of $3$ for $Q_{\rm L}(\omega)$ and $4$ for $Q_{\rm NL}(2\omega)$.}
\label{Fig1}
\end{figure}

The variation of the linear response amplitude $Q_{\rm L}(\omega)$ with amplitude $g$ of the high frequency force demonstrate the resonant behavior for optimal values of $g$ as shown in Fig.~\ref{Fig1}(a). The solid lines represent the linear response amplitude $Q_{\rm L}(\omega)$ calculated from Eq.~(\ref{eq:ana_linearresponse}). These analytical results are compared with the corresponding numerical simulation of $Q_{\rm L}(\omega)$ from Eq.~(\ref{eq-underdamped3}) as shown by the dashed lines in Fig.~\ref{Fig1}(a). The Eq.~(\ref{eq-underdamped3}) is basically the linearized version of the 1D dissipative BJJ where we have retained nonlinear terms upto third order in $z$.  The agreement is found to be fairly satisfactory. To validate our perturbation technique, these two results are compared with the full direct simulation of 1D dissipative BJJ (Eqs.~(\ref{eq:z}) and (\ref{eq:theta})) represented by the dashed dotted lines in Fig.~\ref{Fig1}(a). The direct simulation of the linear response amplitude for full 1D dissipative BJJ also shows resonant behavior of optimal values of $g$ although the resonant peak is shifted a little bit away for optimal values of $g$, with some oscillatory behavior. The oscillatory behavior is possibly due to the effect of nonlinearity in presence of higher strength of the high frequency field. The similar calculations are carried out for three different values of high frequency amplitude $\Omega$, to show similar resonant behavior for different values of $\Omega$. We observe that, in each cases, the peak position is slightly deviated for the full simulation of 1D dissipative BJJ. 

In Fig.~\ref{Fig1}(b), we show the variation of the nonlinear response amplitude $Q_{\rm NL}(2\omega)$ with amplitude $g$ of the high frequency force. The results also demonstrate the resonant behavior for optimal values of $g$. The solid lines represent the nonlinear response amplitude $Q_{\rm NL}(2\omega)$ calculated from Eq.~(\ref{eq:ana_nonlinearresponse}). These analytical results are compared with the numerical simulation of $Q_{\rm NL}(2\omega)$ from Eq.~(\ref{eq-underdamped3}) using Eqs.~(\ref{eq:13}), (\ref{eq:14}) and (\ref{eq:15}), represent by the dashed lines in  Fig.~\ref{Fig1}(b). The agreement is found to be satisfactory. However, we observe the increase in deviation between these two simulations compared with the linear response simulation as shown in Fig.~\ref{Fig1}(a). We also carried out full direct simulation of 1D dissipative BJJ (Eqs.~(\ref{eq:z}) and (\ref{eq:theta})) represent by the dashed-dotted lines in Fig.~\ref{Fig1}(b). The direct simulation of the nonlinear response amplitude for full 1D dissipative BJJ also shows resonant behavior for optimal values of $g$. However, the resonant peak is shifted to large values of $g$ with some oscillatory behavior due to the effect of nonlinearity in presence of higher strength of high frequency field.

\subsection{Zero-phase mode}
We now consider the case of negative onsite interaction in the zero-phase mode with initial conditions $(z_{0},\theta_{0})=(0.1,0)$. The parameters are chosen as $\Lambda_0=-1.2$ and $\gamma=2.4$, corresponding to $\Lambda_0/\gamma=-0.5$. While the other parameters, $\omega$ and $c$ are kept the same as in the $\pi$-phase mode.

\begin{figure}[htbp]
\centering
\includegraphics[width=0.48\textwidth]{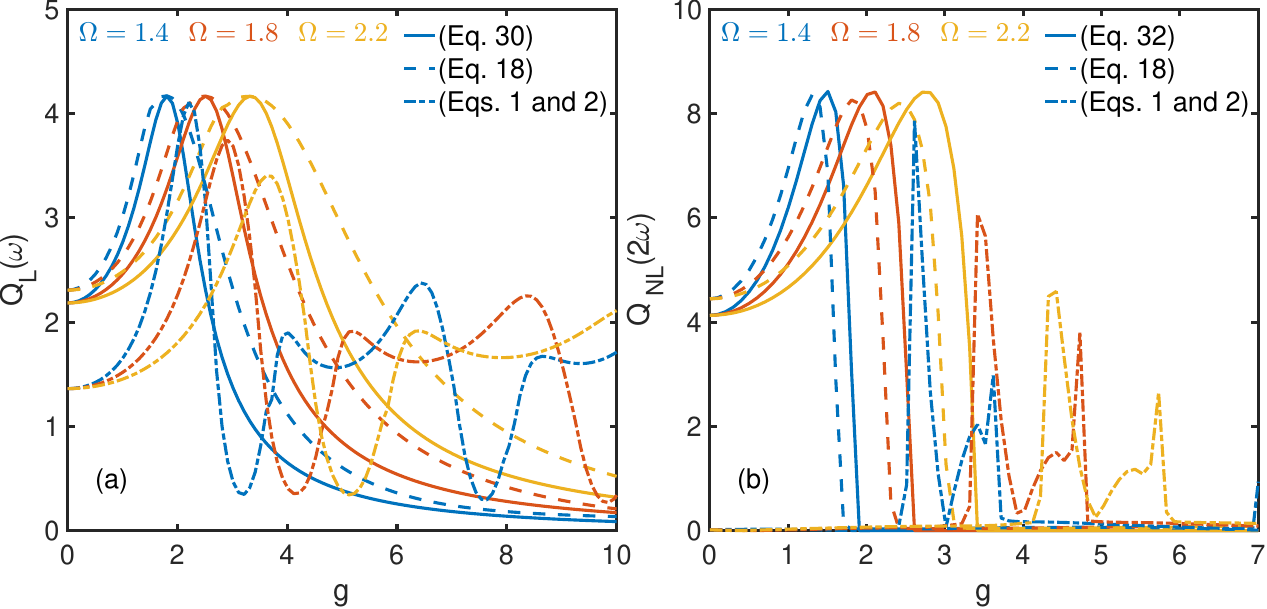}
\caption{(Color Online) (a) Linear response amplitude $Q_{\rm L}(\omega)$ and (b) nonlinear response amplitude $Q_{\rm NL}(2\omega)$ of the 1D dissipative BJJ for $\Lambda_{0}/\gamma = -0.5$, with a negative interaction parameter in the zero-phase mode, in response to a low-frequency signal $c\cos(\omega t)$ under the influence of a high-frequency drive $g\cos(\Omega t)$. Results are shown for three different values of $\Omega = 1.4$ (blue), $1.8$ (orange), and $2.2$ (yellow), with fixed parameters $c = 0.05$ and $\omega = 0.1$. The solid lines represent the analytical results for $Q_{\rm L}$ (Eq.~\ref{eq:ana_linearresponse}) and $Q_{\rm NL}$ (Eq.~\ref{eq:ana_nonlinearresponse}). The dashed lines correspond to numerical simulations of $Q_{\rm L}(\omega)$ and $Q_{\rm NL}(2\omega)$ using Eq.~(\ref{eq-underdamped3}). The dash-dotted lines represent numerical simulations using Eqs.~(\ref{eq:z}) and (\ref{eq:theta}). For better visibility of the resonant behavior as a function of $g$, the results from the full direct simulation of the 1D dissipative BJJ (Eqs.~(\ref{eq:z}) and (\ref{eq:theta})) are multiplied by factors of $3$ for $Q_{\rm L}(\omega)$ and $4$ for $Q_{\rm NL}(2\omega)$.}
\label{Fig2}
\end{figure}

The variation of the linear response amplitude $Q_{\rm L}(\omega)$ and nonlinear response amplitude $Q_{\rm NL}(2\omega)$ as a function of $g$ is shown in Fig.~\ref{Fig2}(a) and Fig.~\ref{Fig2}(b), for three different values of high frequency amplitude $\Omega$. Qualitatively, the behavior remains similar to that observed for $\pi$-phase mode: $Q_{\rm L}(\omega)$ and $Q_{\rm NL}(2\omega)$ exhibits pronounced peaks at optimal values of $g$, and this feature persists for all considered values of $\Omega$. Note that in the zero-phase mode, we used $\alpha_0$ and $\beta_{0}$ in Eqs.~(\ref{eq-underdamped3}), (\ref{eq:ana_linearresponse}) and (\ref{eq:ana_nonlinearresponse}). We find that the overall structure of the response curves, including the presence of well-defined peak and its dependence on $\Omega$, is largely unaffected by the change in the sign of the interaction parameter. The agreement between the analytical (Eqs.~(\ref{eq:ana_linearresponse}) and (\ref{eq:ana_nonlinearresponse})) and numerical results (Eq.~(\ref{eq-underdamped3})) continues to be satisfactory in both linear and nonlinear response cases as shown in Fig.~\ref{Fig2}(a) and Fig.~\ref{Fig2}(b). We also observed the similar deviation of linear and nonlinear response peak position for the full simulation of 1D dissipative BJJ in zero-phase mode. This confirms the validity of the theoretical approach used to describe both the linear and nonlinear response in the presence of dissipation and high frequency force. Overall, these results demonstrate that the resonance persists even in negative interaction regime and the theoretical framework described here capture the essential physics of the 1D dissipative BJJ in presence of perturbation.

\subsection{Dependence of resonance position on system parameters}
We now investigate the dependence of the position of the maximum amplitude $g_{\max}$ on the system parameters. Here, we present the results only for positive interaction parameters (i.e., in the $\pi$-phase mode), since the behavior for negative interactions is qualitatively similar. Additionally, we focus on simulation of $g_{\rm max}$ obtained from linear response $Q_{\rm L}(\omega)$ (Eq.~(\ref{eq:ana_linearresponse})) and nonlinear response $Q_{\rm NL}(2\omega)$ (Eq.~(\ref{eq:ana_nonlinearresponse})) and compared our results with that obtained from Eq.~(\ref{eq-underdamped3}).

\begin{figure}[htbp]
\centering
\includegraphics[width=0.48\textwidth]{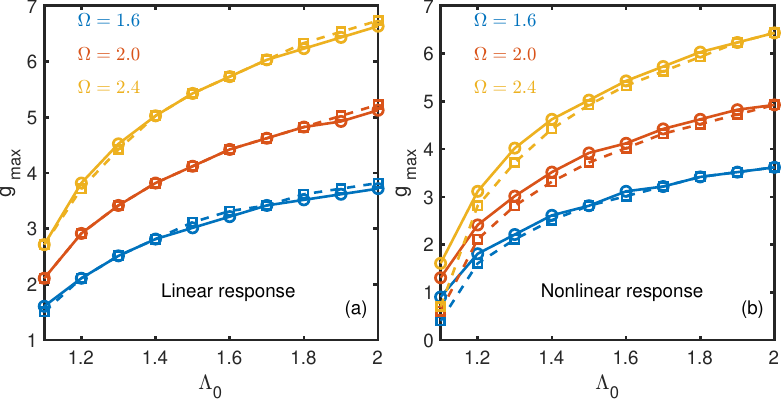}
\caption{(Color Online) The dependence of the position of the maximum amplitude ($g_{\rm max}$) on the interaction parameter $\Lambda_0$ is shown for (a) the linear response $Q_{\rm L}(\omega)$ and (b) the nonlinear response $Q_{\rm NL}(2\omega)$, for three different values of the high-frequency parameter $\Omega =$ $1.6$ (blue), $2.0$ (orange), and $2.4$ (yellow). The solid lines with markers represent the results obtained from Eq.~(\ref{eq:ana_linearresponse}) for $Q_{\rm L}$ and Eq.~ (\ref{eq:ana_nonlinearresponse}) for $Q_{\rm NL}$, while the dashed lines with markers correspond to the results obtained from Eq. (\ref{eq-underdamped3}). Throughout the simulation, we have fixed the dissipation coefficient $\gamma=2.4$ for both linear and nonlinear response cases. In both cases, the other parameters are fixed at $\omega = 0.1$ and $c = 0.05$ throughout the simulations.}
\label{Fig3}
\end{figure}

\emph{Dependence on interaction $\Lambda_{0}$}: To corroborate our analysis, we first examine the variation of $g_{\max}$ with the interaction parameter $\Lambda_0$ for three different values of the high-frequency amplitude $\Omega$. Throughout the simulation, we have fixed the dissipation coefficient $\gamma=2.4$ for both linear and nonlinear response cases. In Fig.~\ref{Fig3}(a), we show the variation of $g_{\max}$ as a function of $\Lambda_0$ for $\Omega = 1.6, 2.0,$ and $2.4$ in the linear response regime. 
The numerical results obtained from Eq.~(\ref{eq-underdamped3}) are compared with the analytical prediction from Eq.~(\ref{eq:ana_linearresponse}), showing good agreement across the entire range. 
We observe that $g_{\max}$ increases monotonically with $\Lambda_0$ for all values of $\Omega$. 
Moreover, increasing $\Omega$ shifts $g_{\max}$ to higher values, indicating a strong dependence on the driving amplitude. 
The overall trend suggests a sublinear (approximately square-root-like) growth at intermediate values of $\Lambda_0$, followed by a tendency toward saturation at larger $\Lambda_0$.

In Fig.~\ref{Fig3}(b), we present the variation of $g_{\rm max}$ with $\Lambda_0$ in the nonlinear response regime. Similar to the linear response case, the numerical results obtained from Eq.~(\ref{eq-underdamped3}) show very good agreement with the analytical prediction from Eq.~(\ref{eq:ana_nonlinearresponse}), confirming the validity of the theoretical model in presence of time-dependent perturbation even beyond the linear regime. As seen in Fig.~\ref{Fig3}(b), $g_{\rm max}$ again increases monotonically with $\Lambda_{0}$ for three different values of $\Omega$. However, compared to Fig.~\ref{Fig3}(a), the growth is more gradual, and the overall magnitude of $g_{\rm max}$ is systematically reduced. The dependence on $\Omega$ remains qualitatively similar, with larger $\Omega$ leading to higher values of $g_{\rm max}$, although the separation between different $\Omega$ curves is slightly less pronounced that in the linear regime.


For small $\Lambda_{0}$, the behavior is qualitatively similar in both the linear and nonlinear response regimes, as shown in Fig.~\ref{Fig3}(a) and Fig.~\ref{Fig3}(b). In both cases, $g_{\rm max}$ increases monotonically with $\Lambda_{0}$ and is systematically enhanced with increasing $\Omega$. A quantitative difference, however, emerges in the weak-interaction regime: in the nonlinear case, the initial slope is noticeably smaller, and $g_{\rm max}$ starts from lower values compared to the linear response. This indicates that nonlinear effects suppress the resonance amplitude at small $\Lambda_{0}$. Despite this difference, the qualitative behavior remains the same in both cases. Overall, the nonlinear response is characterized by a reduced initial growth rate and weaker sensitivity at small $\Lambda_{0}$, highlighting the role of nonlinear corrections in moderating the system dynamics.

\begin{figure}[htbp]
\centering
\includegraphics[width=0.48\textwidth]{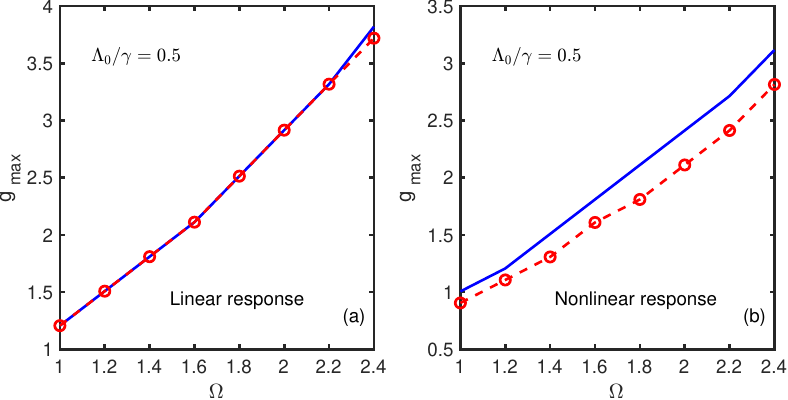}
\caption{(Color Online) The variation of the position of the maximum amplitude $g_{\rm max}$ with the high-frequency parameter $\Omega$ is shown for (a) the linear response $Q_{\rm L}(\omega)$ and (b) the nonlinear response $Q_{\rm NL}(2\omega)$, with $\Lambda_0/\gamma = 0.5$. The blue solid line represents the results obtained from Eq.~(\ref{eq:ana_linearresponse}) for $Q_{\rm L}$ and Eq.~(\ref{eq:ana_nonlinearresponse}) for $Q_{\rm NL}$, while the dashed red line with markers corresponds to the results obtained from Eq.~(\ref{eq-underdamped3}). In both cases, the other parameters are fixed at $\omega = 0.1$ and $c = 0.05$ throughout the simulations.}
\label{Fig4}
\end{figure}

\emph{Dependence on high frequency $\Omega$}:
In Fig.~\ref{Fig4}(a), we present the dependence of $g_{\max}$ on $\Omega$ for a fixed ratio $\Lambda_0/\gamma = 0.5$ in the linear response regime. The solid blue curve corresponds to the analytical results obtained from Eq.~(\ref{eq:ana_linearresponse}), while the dashed red curve represents the results obtained from Eq.~(\ref{eq-underdamped3}). The agreement between the two is excellent. We find that $g_{\max}$ increases approximately linearly with $\Omega$ over the considered range, indicating that the high-frequency drive predominantly controls the amplitude in this regime.

In Fig.~\ref{Fig4}(b), we present the dependence of $g_{\rm max}$ on $\Omega$ in the nonlinear response regime for the same ratio $\Lambda_{0}/\gamma=0.5$. As in Fig.~\ref{Fig4}(a), $g_{\rm max}$ exhibits an overall linear increase with $\Omega$, indicating that the high frequency perturbation continues to predominantly control the resonance position. However, in contrast to the linear response regime, a slight deviation between the numerical results obtained from Eq.~(\ref{eq-underdamped3}) and the analytical prediction from Eq.~ (\ref{eq:ana_nonlinearresponse}) becomes visible, which gradually increase at higher $\Omega$. This reflects the onset of nonlinear effects that are not fully captured by the analytical approximation, while the general scaling behavior remains consistent with Fig.~\ref{Fig4}(a).

\begin{figure}[htbp]
\centering
\includegraphics[width=0.48\textwidth]{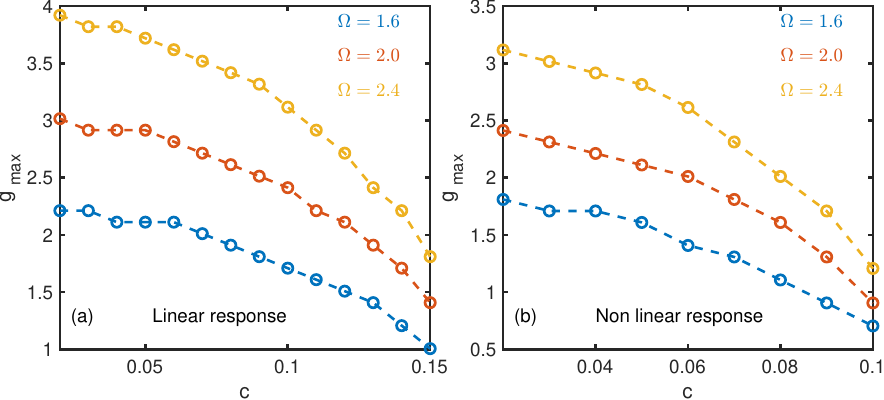}
\caption{(Color Online) The dependence of the position of the maximum amplitude ($g_{\rm max}$) on $c$ is shown for (a) the linear response $Q_{\rm L}(\omega)$ and (b) the nonlinear response $Q_{\rm NL}(2\omega)$, for three different values of the high-frequency parameter $\Omega = 1.6$ (blue), $2.0$ (orange), and $2.4$ (yellow). The dashed lines with markers represent the results obtained from Eq.~(\ref{eq-underdamped3}). Here, we fix $\Lambda_0/\gamma = 0.5$, and the other parameters are set to $\omega = 0.1$ and $c = 0.05$ throughout the simulations.}
\label{Fig5}
\end{figure}

\emph{Dependence on low frequency force $c$}: Finally, in Fig.~\ref{Fig5}(a), we show the variation of $g_{\max}$ as a function of the parameter $c$ for three different values of $\Omega$, at fixed $\Lambda_0/\gamma = 0.5$ in the linear response regime. 
Here, we present only the results obtained from Eq.~(\ref{eq-underdamped3}), as Eq.~(\ref{eq:ana_linearresponse}) shows no additional dependence on $c$. 
We observe that for small values of $c$, $g_{\max}$ remains nearly constant, while for larger values of $c$, it decreases monotonically. This behavior indicates that $c$ plays a suppressing role at higher values, reducing the maximum amplitude.

In Fig.~\ref{Fig5}(b), we present the variation of $g_{\rm max}$ as a function of $c$ in the nonlinear response regime for the same set of high frequencies $\Omega$ and fixed $\Lambda_0/\gamma = 0.5$. Similar to Fig.~\ref{Fig5}(a), $g_{\rm max}$ remains nearly constant for small values of $c$ and decreases monotonically as $c$ increases, confirming that $c$ continues to play a suppressing role even in the nonlinear response regime. However, in contrast to the linear response case, the overall magnitude of $g_{\rm max}$ is reduced, and the decay with $c$ becomes more pronounced, especially at higher $\Omega$. This indicates that nonlinear effects enhance the suppression induced by $c$, leading to a stronger reduction of the maximum amplitude while preserving the same qualitative trend observed in Fig.~\ref{Fig5}(a).

These results demonstrate that $g_{\max}$ is strongly influenced by both the interaction parameter $\Lambda_0$ and the driving amplitude $\Omega$, while its dependence on $c$ is weak for small values and becomes more pronounced as $c$ increases.

\section{Conclusion}\label{sec:5}
In this paper, we have investigated the linear and nonlinear response of a 1D dissipative BJJ subjected simultaneously to a weak low frequency probe and a rapidly oscillating high frequency external drive. Starting from the dissipative two-mode BJJ equations, we derived an effective higher order nonlinear equation for the population imbalance by retaining the leading nonlinear correction upto cubic order. Using the method of time-scale separation and perturbative analysis, we obtained analytical expressions for both the linear response amplitude at the fundamental frequency and nonlinear response amplitude at the second harmonic frequency.

Our analysis demonstrates that the high frequency periodic modulation modifies the effective potential landscape of the 1D dissipative BJJ and gives rise to resonance like enhancement of the response amplitudes for optimal values of the high frequency driving strength. In particular, the interference between the slow and fast time scales breaks the effective symmetry of the potential around the stationary state, generating a finite nonlinear response at the second harmonic frequency. As a consequence, the system exhibits not only enhanced linear response but also pronounced nonlinear response under appropriate conditions.

The analytical predictions for both the linear response $Q_{\rm L}(\omega)$ and nonlinear response $Q_{\rm NL}(2\omega)$ were compared with numerical simulations of the reduced nonlinear equation as well as with direct simulations of the full 1D dissipative BJJ. We found overall satisfactory agreement between analytical and numerical results over a broad range of parameters in both the $\pi$-phase mode with positive interaction and the zero-phase mode with negative interaction. Although small deviations and shifts of the resonance peak appear in the full numerical simulations mainly due to higher order nonlinear effects neglected in the analytical treatment, the essential resonant behavior remains robust.

We further analyzed the dependence of the resonance position $g_{\rm max}$ on the interaction parameter $\Lambda_{0}$, the high frequency parameter $\Omega$, and the low-frequency forcing amplitude $c$. The results show that $g_{\rm max}$ increases monotonically with both $\Lambda_{0}$ and $\Omega$ whereas increasing $c$ suppresses the resonance amplitude at larger values. In the nonlinear response regime, the growth of $g_{\rm max}$ is comparatively weaker, indicating that nonlinear corrections moderate the effective dynamics of the system. These observations confirm that the interplay between dissipation, interaction strength, and external periodic forcing governs the emergence and control of resonance in 1D dissipative BJJ.  

An important outcome of the present study is that the resonance phenomena persists for both positive and negative interaction regimes, indicating the generality and robustness of the mechanism. The effective symmetry breaking induced by the rapidly oscillating field provides a controllable route for manipulating higher harmonic generation and nonlinear response in driven BJJ. Since dissipative BJJ is experimentally realizable in ultra-cold atomic setups \cite{josephson:exp4,Josephson:dissipation,Mennemann:2021,Yuri:2021,Ji:2022}, the present results may be useful for controlling coherent dynamics, population imbalance oscillations, and nonlinear collective modes through external periodic modulation. The present work may also be extended to investigate the role of stronger nonlinearities, quantum fluctuations, many-body effects beyond mean-field approximation, and periodically driven multi-well or lattice bosonic systems \cite{Eckardt:2017,Bukov:2015} which may reveal further non-equilibrium dynamical phenomena in interacting quantum gases.

\section{Acknowledgment}
AKS gratefully acknowledges Deb Shankar Ray for introducing him to the linear and nonlinear response methods and for numerous inspiring discussions and valuable suggestions. The work is supported by the Japan Society for the Promotion of Science (JSPS) Grants-in-Aid for Scientific Research (KAKENHI) Grant No. JP25K00925.

\bibliography{biblio}

\end{document}